\documentclass[runningheads]{llncs}
\usepackage{amsfonts}
\usepackage{array}
\usepackage{booktabs}
\usepackage[T1]{fontenc}
\usepackage{graphicx}
\usepackage{listings}
\usepackage[colorlinks,linkcolor=blue,anchorcolor=blue,citecolor=blue]{hyperref}
\usepackage{multirow}
\newcommand{\ie}{\textit{i.e.}}
\newcommand{\eg}{\textit{e.g.}}
\newcommand{\etal}{\textit{et al.}}

\begin{document}

\title{What Makes for Automatic Reconstruction of Pulmonary Segments}
\titlerunning{What Makes for Automatic Reconstruction of Pulmonary Segments}

\author{Kaiming Kuang\inst{1}\thanks{Equal contributions: Kaiming Kuang and Li Zhang.} \and Li Zhang\inst{1\star} \and Jingyu Li\inst{3} \and Hongwei Li\inst{4} \and Jiajun Chen\inst{1} \and \\Bo Du\inst{3} \and Jiancheng Yang\inst{1,2,5}\thanks{Corresponding author: Jiancheng Yang (jekyll4168@sjtu.edu.cn).}}

\authorrunning{K. Kuang \etal}

\institute{Dianei Technology, Shanghai, China \and 
Shanghai Jiao Tong University, Shanghai, China\\\email{jekyll4168@sjtu.edu.cn} 
\and
Wuhan University, Hubei, China 
\and 
Technical University of Munich, Munich, Germany \and 
EPFL, Lausanne, Switzerland}

\maketitle

\begin{abstract}
3D reconstruction of pulmonary segments plays an important role in surgical treatment planning of lung cancer, which facilitates preservation of pulmonary function and helps ensure low recurrence rates. However, automatic reconstruction of pulmonary segments remains unexplored in the era of deep learning. In this paper, we investigate \textit{what makes for automatic reconstruction of pulmonary segments.} First and foremost, we formulate, clinically and geometrically, the anatomical definitions of pulmonary segments, and propose evaluation metrics adhering to these definitions. Second, we propose ImPulSe (\textbf{Im}plicit \textbf{Pul}monary \textbf{Se}gment), a deep implicit surface model designed for pulmonary segment reconstruction. The automatic reconstruction of pulmonary segments by ImPulSe is accurate in metrics and visually appealing. Compared with canonical segmentation methods, ImPulSe outputs continuous predictions of arbitrary resolutions with higher training efficiency and fewer parameters. Lastly, we experiment with different network inputs to analyze what matters in the task of pulmonary segment reconstruction. Our code is available at \href{https://github.com/M3DV/ImPulSe}{https://github.com/M3DV/ImPulSe}.
\keywords{pulmonary segments \and surface reconstruction \and implicit fields.}
\end{abstract}

\section{Introductions}
Pulmonary segments are anatomical subunits of pulmonary lobes. There are 18 segments in total, with eight in the left lung and ten in the right~\cite{Boyden1945TheIA,Jackson1943CorrelatedAA,Ugalde2007LobesFA}. Unlike pulmonary lobes, pulmonary segments are not defined by visible boundaries but bronchi, arteries and veins. Concretely, pulmonary segments should include their segmental bronchi and arteries while establishing their boundaries along intersegmental veins~\cite{Oizumi2014,Frick2017}. However, boundaries between adjacent segments are ambiguous since segmenting planes are valid as long as they separate segmental bronchi and arteries while lying roughly in the area of intersegmental veins.

Automatic reconstruction of pulmonary segments helps determine the appropriate resection method in surgical treatment of lung cancer. Lobectomy (excising the affected lobe entirely) and segmentectomy (excising only the affected segment) are two major resection methods for early stage lung cancer. As the standard care of early stage lung cancer, lobectomy is challenged by segmentectomy as it preserves more pulmonary function, reduces operation time and blood loss while leading to similar recurrence rates and survival~\cite{Schuchert2007,Frick2017,Wisnivesky2010,Handa2021,Harada2005}. However, segmentectomy should only be considered when surgical margins can be guaranteed to ensure low recurrence rates~\cite{Schuchert2007,Oizumi2011}. Therefore, it is crucially important to reconstruct pulmonary segments before performing pulmonary surgeries.

Recent years have witnessed the great success of deep learning in medical image segmentation~\cite{Ronneberger2015UNetCN,Milletari2016VNetFC,Zhou2020,Isensee2021}, as well as in segmentation of pulmonary structures such as lobes, airways and vessels~\cite{Gerard2019PulmonaryLS,Gerard2019FissureNetAD,Nardelli2018PulmonaryAC,Qin2021LearningTC}. Nonetheless, automatic reconstruction of pulmonary segments remains poorly understood. First, canonical segmentation methods such as U-Net~\cite{Ronneberger2015UNetCN} are not suitable for this task. These methods create large memory footprint with medical images of original resolutions, and output poor segmentation if inputs are downsampled. Second, it is desired that the pulmonary segment reconstruction model can generate outputs at arbitrary resolutions when given only coarse inputs. Lately, deep implicit functions show great promises in representing continuous 3D shapes~\cite{Park2019DeepSDFLC,Mescheder2019OccupancyNL,Chen2019LearningIF,Peng2020ConvolutionalON,Chibane2020,huang2022representation}. The learned implicit function predicts occupancy at continuous locations, thus is capable of reconstructing 3D shapes at arbitrary resolutions. Moreover, implicit fields can be optimized using irregular points randomly sampled from the entire continuous space, which significantly reduces training costs. These characteristics suggest that implicit functions can be of use in the reconstruction of pulmonary segments.

In this paper, we aim to answer this question: \textit{what makes for automatic reconstruction of pulmonary segments?} First and foremost, we give clear and concrete definitions of the anatomy of pulmonary segments (Section~\ref{sec:def}) and propose evaluation metrics that adhere to problem definitions (Section~\ref{sec:metrics}); Next, we present an implicit-function-based model, named ImPulSe (\textbf{Im}plicit \textbf{Pul}monary \textbf{Se}gment), for the pulmonary segment reconstruction task (Section~\ref{sec:network} and~\ref{sec:training}). Implicit fields render ImPulSe with high training and parameter efficiency and the ability to output reconstruction at arbitrary resolutions. Our proposed method uses only half the training time of U-Net~\cite{Ronneberger2015UNetCN} while having better accuracy. ImPulSe achieves Dice score of 84.63\% in overall reconstruction, 86.18\%  and 87.00\% in segmental bronchus and artery segmentation. Last but not least, we investigate what inputs matter for accurate reconstruction of pulmonary segments (Section~\ref{sec:ablations}). Our experiments indicate that bronchi and vessels play important roles in the reconstruction of pulmonary segments.

\section{Methods}

\begin{figure}[tb]
    \centering
    \includegraphics[width=1\textwidth]{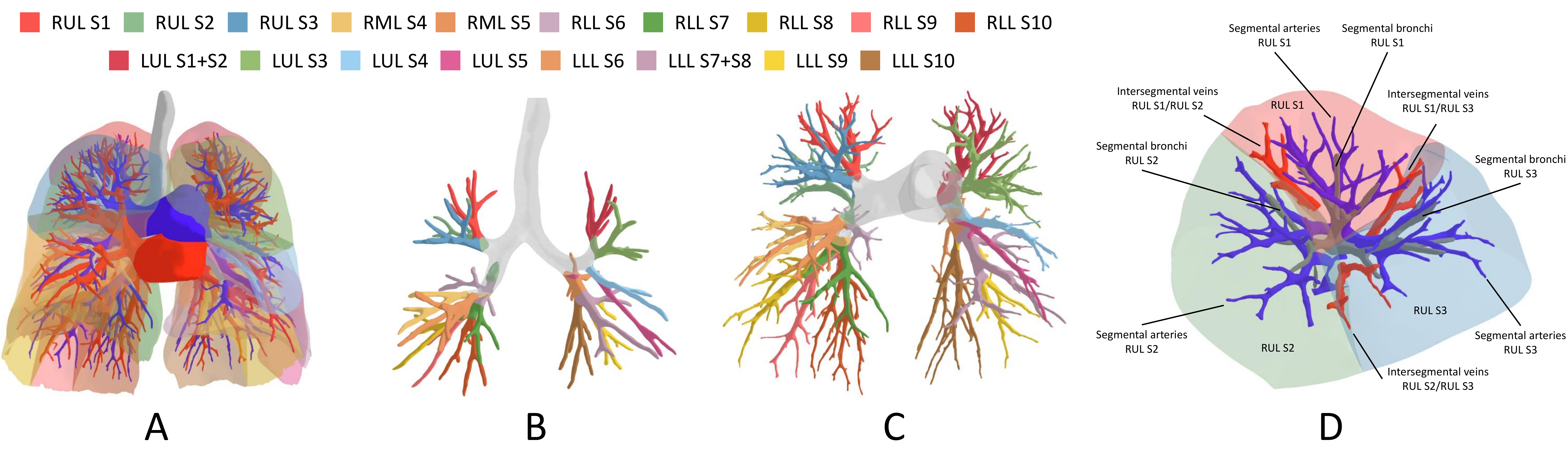}
    \caption{\textbf{Visualization of pulmonary segment anatomy.} \textbf{A}: An overview of pulmonary segments, including bronchi, arteries and veins. \textbf{B, C}: Bronchus and artery tree are separated into segmental groups, each of which occupies a branch of the tree. \textbf{D}: A concrete example of intersegmental boundaries of RUL S1 (middle), RUL S2 (left) and RUL S3 (right). Segmental bronchi, segmental arteries and intersegmental veins are colored in gray, blue and red. Intrasegmental veins are hidden for better visualization. Each segment completely wraps its own segmental bronchi and arteries. Intersegmental boundaries lie on the branch of intersegmental veins.}
    \label{fig:segment_bronchi_artery}
\end{figure}

\subsection{Anatomical Definitions of Pulmonary Segments}\label{sec:def}
To start the investigation of automatic pulmonary segment reconstruction, it is necessary to first set clear anatomical definitions of pulmonary segments. Pulmonary segments are subunits of pulmonary lobes. There are 5 pulmonary lobes: left upper lobe (LUL), left lower lobe (LLL), right upper lobe (RUL), right middle lobe (RML), and right lower lobe (RLL). These lobes are further divided into 18 pulmonary segments, with eight segments in the left lung and ten in the right lung~\cite{Boyden1945TheIA,Jackson1943CorrelatedAA,Ugalde2007LobesFA}. These segments are numbered from S1 to S10 in either left lung and right lung, with two exceptions in the left lung (LUL S1+S2 and LLL S7+S8). Unlike other anatomical structures such as pulmonary lobes, pulmonary segments are not defined by visible boundaries but pulmonary bronchi, arteries and veins. Pulmonary bronchi and vessels (including arteries and veins) expand themselves into tree structures in lungs. Branches of bronchi and vessel trees are then divided into 18 segmental groups (\ie, segmental bronchi and arteries), by which pulmonary segments are defined. Fig.~\ref{fig:segment_bronchi_artery}B and C show the bronchus tree and the artery tree, with segmental groups marked in different colors. Each color represents a segmental bronchus/artery. Concretely, pulmonary segments should satisfy the following three rules~\cite{Oizumi2014,Frick2017}:

\textit{(1) Includes its segmental bronchi.} The volume of a certain segment should completely includes its corresponding segmental bronchus, as in Fig.~\ref{fig:segment_bronchi_artery}B.

\textit{(2) Includes its segmental arteries.} A certain segment should contain its segmental artery as in the case of segmental bronchi. See Fig.~\ref{fig:segment_bronchi_artery}C.

\textit{(3) Establishes its boundaries along intersegmental veins.} Pulmonary veins can be subdivided into intrasegmental or intersegmental veins. Intersegmental veins serve as important landmarks indicating boundaries between adjacent segments. Specifically, intersegmental planes should follow intersegmental veins.

Fig.~\ref{fig:segment_bronchi_artery}D gives a concrete example of pulmonary segment boundaries (RUL S1 in the left, RUL S2 in the middle, and RUL S3 in the right). Segmental bronchi, segmental arteries and intersegmental veins are marked in gray, blue and red, respectively. Intrasegmental veins are hidden for better visualization. Each segment completely wraps its own segmental bronchi and arteries. Boundaries between segments lie on the branch of intersegmental veins, as per definitions. Segmental bronchi and arteries do not cross segmental boundaries.

\subsection{Evaluation Metrics}\label{sec:metrics}
Since the automatic reconstruction of pulmonary segments have not been systematically investigated, it is equally important to select appropriate evaluation metrics for this task. Dice score is a commonly-used metric in segmentation,
\begin{equation}
    Dice_{\mathbf{s}}=\frac{2\|\mathbf{Y}_{\mathbf{s}}\cap\hat{\mathbf{Y}}_{\mathbf{s}}\|}{\|\mathbf{Y}_{\mathbf{s}}\|+\|\hat{\mathbf{Y}}_{\mathbf{s}}\|},
\end{equation}
where $\mathbf{Y}_{\mathbf{s}}$ and $\hat{\mathbf{Y}}_{\mathbf{s}}$ are ground-truth and prediction segmentation of points in the set $\mathbf{s}$, and $\|\cdot\|$ denotes the number of elements. Definitions of pulmonary segments suggest that their boundaries are ambiguous. Therefore, it is more important to guarantee accurate reconstruction of segmental bronchi and arteries than segment reconstruction itself. Given that, we evaluate our proposed method and its counterparts not only using the overall Dice score on 18 pulmonary segments ($Dice_{\mathbf{o}}$) but also Dice on segmental bronchi ($Dice_{\mathbf{b}}$) and arteries ($Dice_{\mathbf{a}}$). Note that $Dice_{\mathbf{b}}$ and $Dice_{\mathbf{a}}$ are not binary Dice scores of bronchi and arteries. Instead, they are 19-class pulmonary segment Dice scores evaluated only on voxels of bronchi and arteries. These two metrics measure how accurate bronchi and arteries are divided into their corresponding pulmonary segments (as in Fig.~\ref{fig:segment_bronchi_artery} B and C). We argue that Dice scores of segmental bronchi and arteries are more important metrics than the overall Dice score. We leave veins out of this question because the segment classification of veins, especially intersegmental veins, is ambiguous. We find that Dice scores on veins are consistently lower than bronchi and arteries across all models (see supplementary materials).

\subsection{The ImPulSe Network}\label{sec:network}
\begin{figure}[tb]
    \centering
    \includegraphics[width=0.9\textwidth]{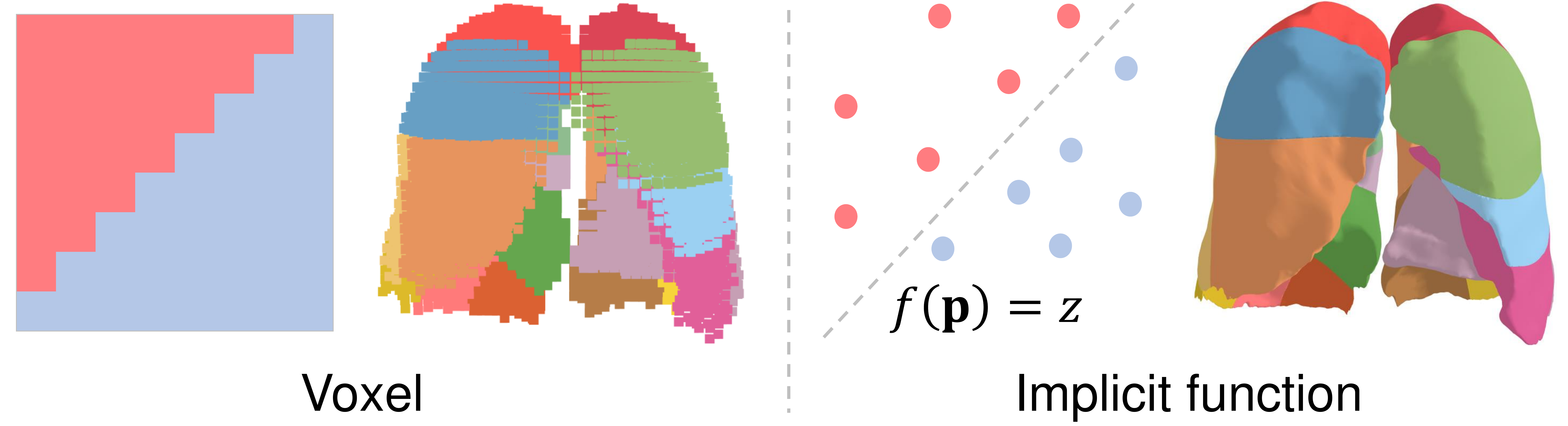}
    \caption{\textbf{Visualization of shape representations with voxel and implicit functions.} Rather than classifying each voxel, implicit functions representing shapes by encoding 3D iso-surfaces with function $f(\mathbf{p})=z$. When queried with a continuous location $\mathbf{p}$, it outputs occupancy $z$ at the location, Thus, implicit functions can output continuous reconstruction with high resolutions and fine details. Please note that this is a conceptual visualization, not a rigorous comparison.}
    \label{fig:voxel_and_implicit}
\end{figure}

\begin{figure}[tb]
    \centering
    \includegraphics[width=0.9\textwidth]{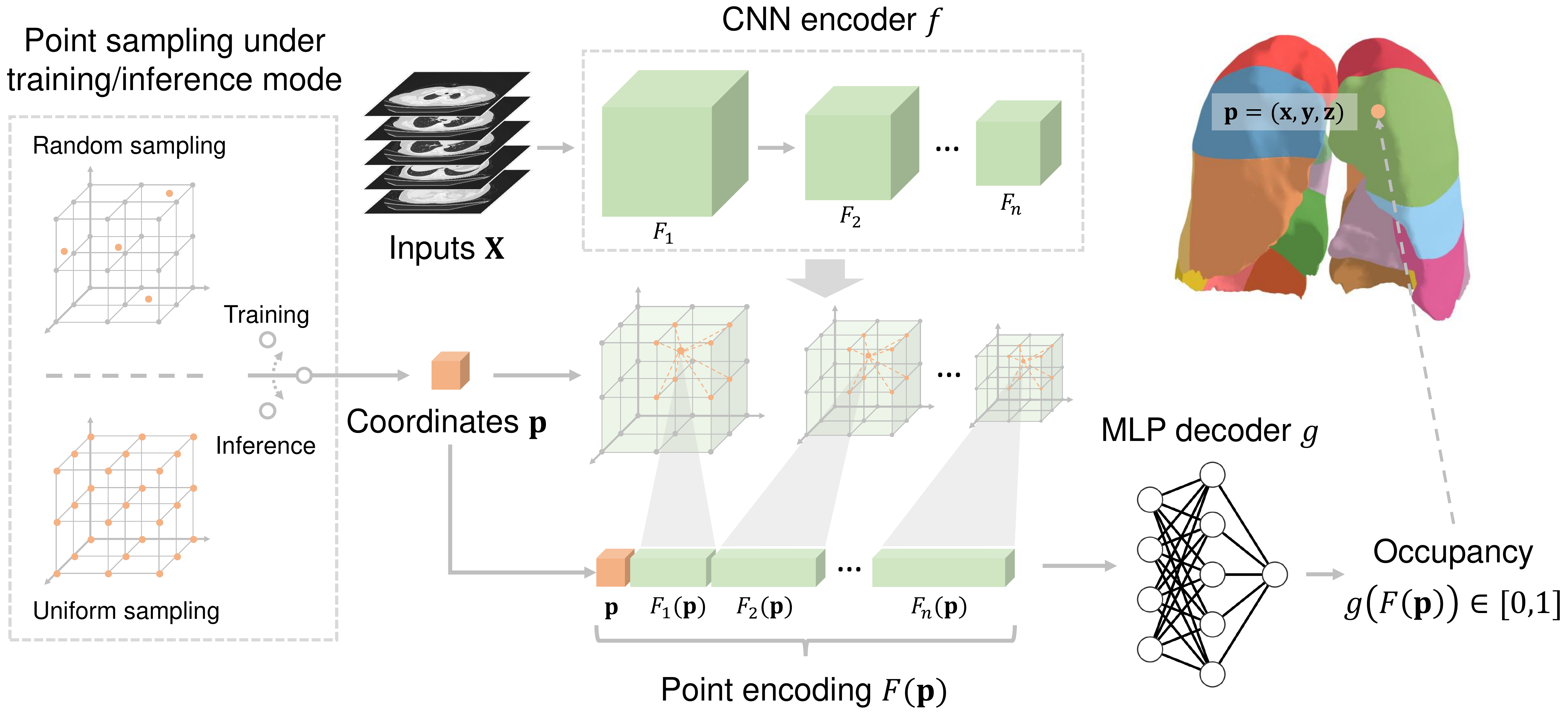}
    \caption{\textbf{The architecture of ImPulSe.} ImPulSe consists of an encoder and a decoder. The encoder $f$ takes a voxel grid $\mathbf{X}$ as input and outputs a feature pyramid ${\mathbf{F}}_{1},{\mathbf{F}}_{2},...,{\mathbf{F}}_{n}$. Features at continuous coordinates $\mathbf{p}$, \ie, ${\mathbf{F}}_{1}(\mathbf{p}),{\mathbf{F}}_{2}(\mathbf{p}),...,{\mathbf{F}}_{n}(\mathbf{p})$, are sampled with tri-linear interpolation and then concatenated with $\mathbf{p}$ to form point encodings $\mathbf{F}(\mathbf{p})$. The decoder $g$ takes $\mathbf{F}(\mathbf{p})$ and predicts occupancy at $\mathbf{p}$. Note that ImPulSe samples far fewer points during training than inference.}
    \label{fig:network}
\end{figure}

To apply deep learning in automatic pulmonary segment reconstruction, one problem shows up: CT images come in different sizes. While high-resolution 3D inputs are not feasible due to large memory footprint, canonical segmentation methods such as U-Net~\cite{Ronneberger2015UNetCN} generate coarse segmentations of fixed sizes given low-resolution inputs (\ie, low-resolution in, low-resolution out), which makes them unsuitable for this task. Unlike U-Net, implicit functions generate continuous outputs of arbitrary resolutions even when only low-resolution inputs are given. Fig.~\ref{fig:voxel_and_implicit} visualizes implicit functions and voxel-based representations. Rather than predicting segmentation in a fixed-size voxel grid, the learned implicit function encodes 3D iso-surfaces with function $f(\mathbf{p})=z$ and predicts occupancy $z$ at a continuous location $\mathbf{p}$, thus is capable of reconstructing 3D shapes at arbitrary resolutions. To take advantage of this in reconstruction of pulmonary segments, we propose ImPulSe (\textbf{Im}plicit \textbf{Pul}monary \textbf{Se}gment), an implicit-function-based model capable of generating reconstruction of arbitrary sizes. The overall architecture of ImPulSe is shown in Fig.~\ref{fig:network}. It consists of an encoder and a decoder. The encoder $f$ is a small CNN (3D ResNet18~\cite{He2016,Tran2018ACL} in our experiments), which takes 3D voxel grids $\mathbf{X}\in\mathbb{R}^{c\times d\times h\times w}$ as inputs and extracts a feature pyramid ${\mathbf{F}}_{1},{\mathbf{F}}_{2},...,{\mathbf{F}}_{n}$. During the decoding stage, 3D continuous coordinates $\mathbf{p}\in{[-1,1]}^{3}$ are sampled from the entire space. This is different than traditional segmentation methods, where the model runs on discrete voxel grids. These locations are queried using tri-linear interpolation to extract features at each coordinate ${\mathbf{F}}_{1}(\mathbf{p}),{\mathbf{F}}_{2}(\mathbf{p}),...,{\mathbf{F}}_{n}(\mathbf{p})$. These features are concatenated with point coordinates $\mathbf{p}$ to form point encodings $\mathbf{F}(\mathbf{p})$, which are then fed into the decoder $g$, a two-layer MLP predicting segment occupancy at each location.

Compared with fully-convolutional methods~\cite{Shelhamer2017FullyCN,Ronneberger2015UNetCN,Chen2017RethinkingAC}, the ImPulSe architecture has two major advantages: it delivers continuous outputs with arbitrarily high resolutions and fine-grained details, while significantly reducing the number of parameters in its decoder. More importantly, thanks to implicit fields, ImPulSe can be trained on continuous locations instead of discrete locations on the voxel grid. These advantages render ImPulSe with high training efficiency while achieving better accuracy than traditional segmentation networks.

\subsection{Training and Inference}\label{sec:training}
During training and inference, ImPulSe alternates between random sampling and uniform sampling (Fig.~\ref{fig:network}). During the training phase, we train ImPulSe on randomly sampled points $\mathbf{p}\in{[-1,1]}^{3}$. Random and continuous sampling implicitly imposes data augmentation and alleviates overfitting while ensuring sufficient coverage of the entire space. Furthermore, it leads to a higher training efficiency compared against canonical segmentation methods, which trains on the entire voxel grid. With random sampling, we are able to train ImPulSe with far fewer points in each batch (\eg, $16^3$ for ImPulSe versus $64^3$ or $128^3$ for its counterparts in our experiments) while achieving better performances. Ground-truth labels on these continuous positions are queried using nearest-neighbor interpolation. The ImPulSe network is trained using a weighted combination of cross-entropy loss and Dice loss~\cite{Milletari2016VNetFC}. All models are trained on 4 NVIDIA RTX 3090 GPUs with PyTorch 1.10.1~\cite{Paszke2019PyTorchAI}. During inference, we replace the random sampling with uniform sampling on the voxel grid, thus output pulmonary segment reconstruction at the original resolution of the CT image.

\section{Experiments}

\subsection{Datasets}
In this study, we compile a dataset containing 800 CT scans with annotations of pulmonary segments, pulmonary bronchi, arteries and veins. To allow elaborative analysis of reconstruction performances, pulmonary veins are further annotated as intrasegmental and intersegmental veins. CT scans are collected from multiple medical centers to improve the generalization performance of our model. Z-direction spacings of these scans range from $0.5mm$ to $1.5mm$. Annotations are manually made by a junior radiologist and confirmed by a senior radiologist. All CT scans are divided into training, validation and test subsets with a 7:1:2 ratio. Results are tuned using the validation set and reported on the test set.

\subsection{Reconstruction Performances}\label{sec:perf}

\begin{table}[tb]
    \centering
        \caption{\textbf{Pulmonary segment reconstruction performances of ImPulSe and its counterparts.} All methods are evaluated on Dice score (\%) of pulmonary segments ($Dice_{\mathbf{o}}$), segmental bronchi ($Dice_{\mathbf{b}}$) and segmental arteries ($Dice_{\mathbf{a}}$). Best metrics are highlighted in bold. With fewer parameters and less training time, ImPulSe achieves better performances than its counterparts. Tr-res and Tr-time denote the input resolution during training and total training time (same number of epochs across models). DNC stands for "did not converge".}
    \label{tab:perf}
    \begin{tabular*}{\linewidth}{@{\extracolsep{\fill}}l>{\centering}p{30pt}>{\centering}p{35pt}>{\centering}p{35pt}>{\centering}p{35pt}cc}
        \toprule
        Methods & Tr-Res & $Dice_{\mathbf{o}}$ & $Dice_{\mathbf{b}}$ & $Dice_{\mathbf{a}}$ & \#Param & Tr-Time \\
        \midrule
        Rikxoort et al.~\cite{Rikxoort2009AutomaticSO} & $128^3$ & 60.63 & 66.01 & 67.54 & 38.9M & 11,009s\\
        FCN~\cite{Shelhamer2017FullyCN} & $128^3$ & 81.70 & 83.67 & 84.92 & 33.4M & 10,746s\\
        U-Net~\cite{Ronneberger2015UNetCN} & $128^3$ & 83.25 & 84.71 & 86.07 & 38.9M & 11,009s \\
        DeepLabv3~\cite{Chen2017RethinkingAC} & $128^3$ & 82.44 & 84.48 & 85.64 & 44.4M & 11,220s \\
        Sliding-Window & $64^3$ & 0.11 & 0.05 & 0.02 & 38.9M & DNC \\
        ImPulSe & $16^3$ & \textbf{83.54} & \textbf{85.14} & \textbf{86.26} & \textbf{33.3M} & \textbf{4,962s} \\
        \bottomrule
    \end{tabular*}
\end{table}

\begin{figure}[tb]
    \centering
    \includegraphics[width=\textwidth]{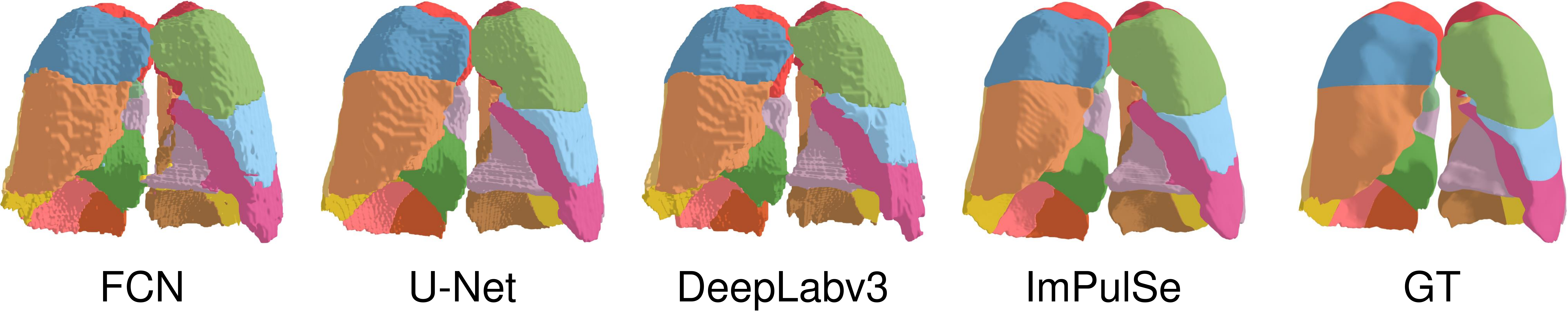}
    \caption{\textbf{Predictions of ImPulSe and some of its counterparts.} Thanks to implicit functions, ImPulSe directly generates reconstruction of original resolution and therefore outputs predictions with smoother surfaces compared with its counterparts.}
    \label{fig:predictions}
\end{figure}

We compare ImPulse and its counterparts, including FCN~\cite{Shelhamer2017FullyCN}, U-Net~\cite{Ronneberger2015UNetCN} and DeepLabv3~\cite{Chen2017RethinkingAC} in the pulmonary segment reconstruction task. Additionally, we roughly reproduce the pulmonary segment reconstruction method in Rikxoort et al.~\cite{Rikxoort2009AutomaticSO} by feeding relative coordinates into a U-Net~\cite{Ronneberger2015UNetCN}. Finally, we train an extra U-Net counterpart using the sliding-window strategy applied on raw resolution image. All models share the same encoder architecture (ResNet18~\cite{He2016,Tran2018ACL}), loss and training schedule to ensure fair comparison. Only CT images are fed as inputs in this experiment. Images are cropped around the lung area and downsampled to the size of $128^3$ in pre-processing except the sliding-window experiment. During inference, ImPulSe directly outputs reconstructions of the original resolution, while predictions of all counterparts (except sliding-window) are upsampled to the original resolution with tri-linear interpolation. Table~\ref{tab:perf} shows performances of ImPulSe and its counterparts. ImPulSe achieves higher performances than its counterparts in all metrics with 14\% fewer parameters and 55\% less training time (compared against U-Net). The light-weighted decoder of ImPulSe (a two-layer MLP) leads to its parameter efficiency, while implicit functions enable training on a small portion of points rather than the entire voxel grid ($16^3$ for ImPulSe versus $64^3$ or $128^3$ for its counterparts) and largely save training time. Sliding-window U-Net does not even converge. This is because learning pulmonary segments requires global information, which is missing in sliding windows. Plus, sliding-window creates extreme class imbalance since only a few segments are present in each window. Fig.~\ref{fig:predictions} 
shows predictions of ImPulSe and some of its counterparts. The reconstruction of ImPulSe has smoother surfaces since it directly outputs prediction at the original resolution.

\subsection{What Makes for Pulmonary Segment Reconstruction?}\label{sec:ablations}

In this section, we evaluate effects of different inputs in reconstruction of pulmonary segments. Various combinations of the following five inputs are considered: CT images (\textit{I}), bronchi (\textit{B}), arteries (\textit{A}), veins (\textit{V}) and lobes (\textit{L}). To obtain these inputs for inference, we train separate models for bronchus, vessel and lobe segmentation. These models are trained using the same dataset and splits as ImPulSe. For all input combinations containing \textit{L}, \textit{B}, \textit{A} and \textit{V}, ImPulSe is trained using ground-truths as input and evaluated using both ground-truths and predictions of aforementioned segmentation models. Table~\ref{tab:inputs} shows performances of ImPulSe given different inputs. Metrics in parentheses are evaluated using ground-truth of \textit{L}, \textit{B}, \textit{A} and \textit{V} as input. Our findings are listed as follows:

\begin{table}[tb]
    \centering
    \caption{\textbf{Reconstruction performances of ImPulSe given different combinations of inputs.} \textit{I}, \textit{L}, \textit{B}, \textit{A}, and \textit{V} represent images, lobes, bronchi, arteries and veins, respectively. We evaluate ImPulSe on both ground-truth and predictions of \textit{L}, \textit{B}, \textit{A} and \textit{V}. Metrics outside/inside parentheses are evaluated on predictions/ground-truth (not applicable for \textit{I}). Best metrics on predictions are highlighted in bold.}
    \label{tab:inputs}
    \begin{tabular}{l>{\centering}p{60pt}>{\centering}p{60pt}>{\centering}p{60pt}>{\centering}p{60pt}>{\centering\arraybackslash}p{60pt}}
        \toprule
        Inputs & \textit{L} & \textit{BAV} & \textit{LBAV} & \textit{I} & \textit{IBAV} \\
        \midrule
        $Dice_{\mathbf{o}}$ & 4.03 (73.77) & 81.61 (81.55) & 72.25 (83.21) & 83.54 (n/a) & \textbf{84.63} (84.88) \\
        $Dice_{\mathbf{b}}$ & 6.11 (79.01) & 85.72 (85.65) & 84.83 (86.77) & 85.14 (n/a) & \textbf{86.18} (86.08) \\
        $Dice_{\mathbf{a}}$ & 6.01 (80.62) & 86.61 (86.75) & 85.53 (88.02) & 86.26 (n/a) & \textbf{87.00} (87.15) \\
        \bottomrule
    \end{tabular}
\end{table}
\paragraph{Bronchi and vessels play important roles.}
Even though \textit{I} outperforms \textit{BAV} in overall Dice score, \textit{BAV} surpasses \textit{I} in bronchi and arteries. Since $Dice_{\mathbf{b}}$ and $Dice_{\mathbf{a}}$ are more important in pulmonary segment reconstruction than the overall Dice, it is clear that bronchi and vessels play crucial roles in this task. Plus, bronchi and vessels are less noisy than images since they are binary while images are encoded in HU values of wide ranges.
\paragraph{Images and BAV are complementary to each other.}
\textit{IBAV} achieves the highest performances in all metrics. We hypothesize that this is due to the complementary effect of \textit{I} and \textit{BAV}. \textit{I} signals some basic knowledge not presented in \textit{BAV}, \eg, contours of lungs. \textit{BAV} highlights bronchi and vessels, which have to be implicitly learned if only image inputs are given.
\paragraph{Lobe as input is prone to overfitting.}
Adding lobes into inputs largely impairs performances (over 9\% overall Dice drop from \textit{BAV} to \textit{LBAV}). We attribute this to the over-simplified structure of pulmonary lobes, which makes neural networks prone to overfitting. This is backed up by the fact that ImPulse only shows large accuracy drops transferring from ground-truth to prediction when it takes lobes as (part of) inputs. Furthermore, lobes include no information related to segmental boundaries, making it difficult to subdivide lobes into segments.

\section{Conclusions}
In this paper, we investigate this question: \textit{what makes for automatic reconstruction of pulmonary segments?} We first set concrete definitions of pulmonary segment anatomy, and suggest approapraite metrics for the reconstruction task. We propose ImPulSe, an implicit-function-based model for pulmonary segment reconstruction. Using implicit fields, ImPulSe generates outputs of arbitrary resolutions with higher training and parameter efficiency compared with its counterparts. In future works, we will investigate if template deformation~\cite{Groueix20183DCODED3C,Wang2020NeuralCF,Jack2018LearningFD,yang2022implicitatlas} and including intersegmental veins are beneficial in this task.

\subsubsection{Acknowledgment.} This work was supported in part by National Science Foundation of China (82071990, 61976238). This work was also supported in part by a Swiss National Science Foundation grant. We would like to thank Lei Liang for his generous help in proofreading, and the anonymous (meta-)reviewers for their valuable comments.

%
%
%
\bibliographystyle{splncs04}
\bibliography{journals_and_conferences,references}

\clearpage

\appendix

\setcounter{table}{0}
\renewcommand{\thetable}{A\arabic{table}}

\setcounter{figure}{0}
\renewcommand{\thefigure}{A\arabic{figure}}

\begin{figure}[h]
    \centering
    \includegraphics[width=1\textwidth]{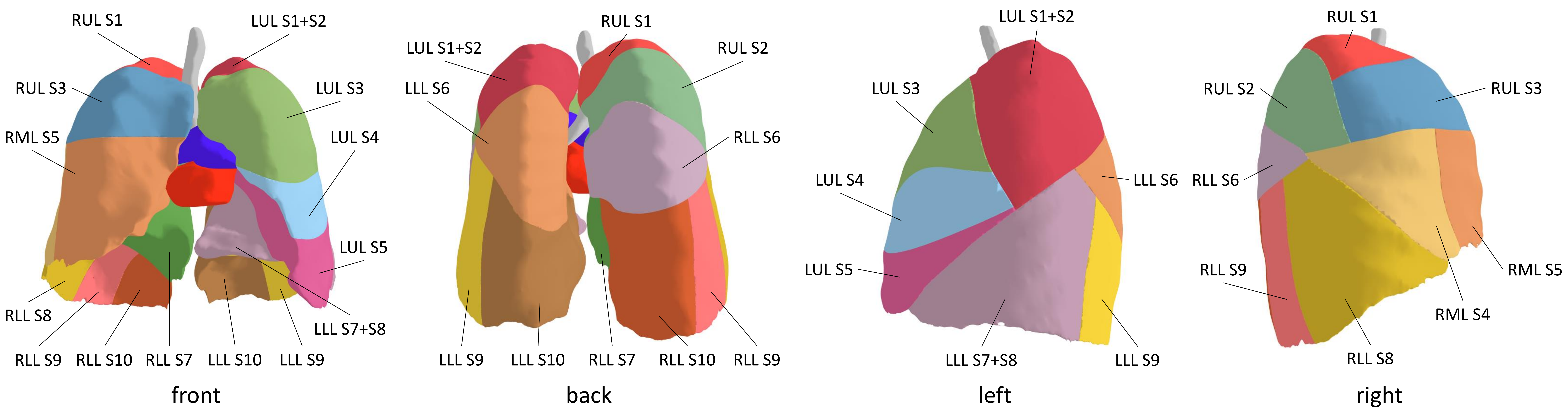}
    \caption{\textbf{The overview of pulmonary segment anatomy.} There are 18 pulmonary segments in total, with eight segments in the left lung and right segments in the right lung. Bronchi, arteries and veins between left and right lung are highlighted in gray, blue and red, respectively.}
\end{figure}

\begin{table}[th]
    \caption{\textbf{Dice scores of bronchi, arteries and veins.} Here veins are further divided into intersegmental veins ($Dice_{\mathbf{inter}}$) and intrasegmental veins ($Dice_{\mathbf{intra}}$). All vein-related metrics, especially $Dice_{\mathbf{inter}}$, are significantly lower than those of bronchi and arteries. This supports our argument that veins are ambiguous in pulmonary segment reconstruction and should not be considered in evaluation. Note that the sliding-window U-Net is omitted in this table since it did not converge.}
    \centering
    \begin{tabular}{l>{\centering}p{40pt}>{\centering}p{40pt}>{\centering}p{40pt}>{\centering}p{40pt}>{\centering\arraybackslash}p{40pt}}
        \toprule
        Methods & $Dice_{\mathbf{b}}$ & $Dice_{\mathbf{a}}$ & $Dice_{\mathbf{v}}$ & $Dice_{\mathbf{inter}}$ & $Dice_{\mathbf{intra}}$ \\
        \midrule
        Rikxoort \etal~\cite{Rikxoort2009AutomaticSO} & 66.01 & 67.54 & 63.01 & 45.92 & 65.42 \\
        FCN~\cite{Shelhamer2017FullyCN} & 83.67 & 84.92 & 77.77 & 61.63 & 80.57 \\
        U-Net~\cite{Ronneberger2015UNetCN} & 84.71 & 86.07 & \textbf{79.53} & \textbf{64.07} & 82.12 \\
        DeepLabv3~\cite{Chen2017RethinkingAC} & 84.48 & 85.64 & 78.57 & 61.65 & 81.49 \\
        ImPulSe & \textbf{85.14} & \textbf{86.26} & 79.51 & 63.80 & \textbf{82.20} \\
        \bottomrule
    \end{tabular}
\end{table}

\end{document}